\documentclass[%
reprint,
superscriptaddress,
amsmath,amssymb,
balancelastpage,
flushbottom,
pra,
floatfix,
]{revtex4-2}

\usepackage{ulem}
\usepackage{dcolumn}% Align table columns on decimal point
\usepackage{bm}% bold math

\usepackage{lmodern}
\usepackage[T1]{fontenc}
\usepackage[utf8]{inputenc}
\usepackage[english]{babel}
\usepackage[dvipsnames]{xcolor}
\usepackage[pdftex]{graphicx}
\usepackage{multirow}
\usepackage[colorlinks,allcolors=blue]{hyperref}

\renewcommand{\d}{\mathrm d}
\renewcommand{\phi}{\varphi}

\begin{document}
\author{A.\ A.\ Fomin} 
\author{G.\ G.\ Kozlov}
\author{M.~Yu.\ Petrov}
\affiliation{Spin Optics Laboratory, St.~Petersburg State University, 198504 St.~Petersburg, Russia}

\author{D.~S.\ Smirnov}
\affiliation{Ioffe Institute, 194021 St.~Petersburg, Russia}

\author{M.~V.\ Petrenko}
\affiliation{Ioffe Institute, 194021 St.~Petersburg, Russia}
 
\author{V.~S.\ Zapasskii}
\affiliation{Spin Optics Laboratory, St.~Petersburg State University, 198504 St.~Petersburg, Russia}

\begin{abstract} 
For isotropic media, the magnetic field applied across the light propagation direction affects their optical properties only in the second order, and its effect proves to be much weaker than in the longitudinal field. In this work, we show that, under resonant excitation well beyond the linear regime, the situation changes drastically: A small magnetic linear anisotropy considerably increases, and an even stronger new quadrupole dichroism emerges. The latter manifests itself as the $90^\circ$-periodic azimuthal dependence of the transmission and anisotropic rotation of the probe polarization plane. These effects are described microscopically in a toy model, and their symmetry analysis is presented. Both are observed experimentally on the $D2$ resonance of cesium vapor and agree with the theoretical predictions. The large magnitude of the quadrupole dichroism makes it promising for magnetometric applications and for studying the effects of anisotropic bleaching.\looseness=-2
\end{abstract}

\title {Magnetic quadrupole dichroism in isotropic medium}
\maketitle

\section{Introduction}
\vspace{-0.5em}
Alkali-metal vapors characterized by strong optical transitions and long relaxation times over the ground-state sublevels are known to be highly susceptible to the action of optical fields and thus appear to be highly suitable model objects of nonlinear optics ~\cite{happer,budker1}. 
Under resonant laser excitation, the effects of light-induced anisotropy are easily observed in these systems even at relatively low levels of the light power densities~\cite{nonlinear1,nonlinear2,nonlinear3}. 
The nonlinear response of the atomic system results from redistribution of populations over the quantum-state levels (such as optical orientation and optical alignment)~\cite{skrot,anderson} or may be related to the effects of optically induced coherence in multilevel structures (such as electromagnetically induced transparency~\cite{EIT1,EIT3,EIT2}  or coherent population trapping~\cite{CPT1,CPT2}).
%\looseness=-1

The symmetry of the light-induced anisotropy, according to the Curie's principle, is usually controlled by the symmetry of the optical field: in the field of linearly or circularly polarized light, the medium acquires linear or circular anisotropy, respectively. 
The effects of linear magneto-optics (the effects of magnetic-field-induced anisotropy) have an apparent analogy with the above effects of the light-induced anisotropy. Specifically, in the longitudinal magnetic field (in the so-called {\it Faraday geometry}), the medium exhibits the effects of circular anisotropy (the Faraday effect and magnetic circular dichroism), while, in the transverse field (in the {\it Voigt geometry}), it exhibits effects of linear anisotropy (magnetic linear birefringence and magnetic linear dichroism). 
The effects of the latter type, in small magnetic fields, are quadratic versus field strength and, usually, are much smaller than the linear (versus field) effects of circular anisotropy. 
%\looseness=-1

However, this conclusion ceases to be valid when the medium is noticeably affected by the probe light. 
Generally, the nonlinear response has a richer symmetry and can reveal anisotropy, which is hidden in the linear response~\cite{Bloembergen,Boyd}. This is well known, for example, for four-wave mixing~\cite{4wm1,4wm2,4wm3}, second harmonic generation~\cite{2h1,2h2,2h3}, and the photogalvanic effect~\cite{PGE1,PGE2,PGE3} also known as the nonlinear Hall effect~\cite{NHE2,NHE3}. 
%\looseness=-1

In this paper, we bring attention to strong and very specific nonlinear magneto-optical effects in isotropic media, e.g., in alkali-metal vapors under resonant excitation, in the Voigt geometry. 
We show that these effects, stemming from anisotropic bleaching of the atomic system, are revealed as anisotropic (angle-dependent) polarization plane (PP) rotation and as a specific linear dichroism with the $90^\circ$ periodicity in transmission of a linearly polarized light. 
Such a nonlinear effect of linear dichroism, which is tentatively called here the {\it magnetic quadrupole dichroism}, is forbidden in linear optics. 
In principle, it can be observed in any anisotropic saturable absorber. 
However, we consider here only the case of an atomic medium in a magnetic field. 
In this case, the anisotropy of the bleaching is determined by the difference in optical alignment in two distinguished directions: along and across the field.
%\looseness=-1

The paper is organized as follows. In Sec.~\ref{sec:II}, we present a simple model of anisotropic saturable absorber capable of describing nonlinear effects of transmission and PP rotation in alkali-metal vapor in a strong resonant optical field. 
In this model, the $\pi$ and $\sigma$ transitions of the atomic system are identified with mutually orthogonal linear oscillators of the model saturable absorber. 
The conditions to be met for the observation of the quadrupole dichroism are formulated. 
In Sec.~\ref{sec:III}, we {perform a general symmetry analysis of the above nonlinear magneto-optical effects}. 
In Sec.~\ref{sec:IV}, we present the experimental evidence for the quadrupole dichroism on the $D2$ line of cesium under different experimental conditions. 
In Sec.~\ref{sec:V}, we discuss the results and the possibility of their use both in fundamental research and in applications.  
%\looseness=-1

\section{Simple model of the nonlinear dichroism}
\label{sec:II}
\subsection{General}
\vspace{-0.5em}
When treating this problem, we proceed from the premise that the magnetic field applied to an isotropic medium across the light beam creates two distinguished directions (along and across the field) that determine the field-induced anisotropy of the medium and specify its normal polarization modes~\cite{nye}.   
These are the basic considerations that should be used to qualitatively describe the effects of the polarized light transformation in atomic vapor (or another anisotropic saturable absorber).  

We assume that total effect of saturable absorption of the medium, in our case of atomic system, is its independent bleaching over the two eigenmodes corresponding to the $\pi$ and $\sigma$ components of the optical transition (with $\pi$\ polarization aligned along the applied magnetic field). 
The light-induced anisotropy of the medium under consideration arises not only because of different laws of bleaching in two eigenmodes but, primarily, from the fact of nonlinearity proper: in the framework of nonlinear optics, equality of optical characteristics of the medium in the two distinguished directions does not make the medium isotropic. 
This is an essential distinction between the effects of linear anisotropy in linear and nonlinear optics. 

The following treatment of the evolution of linearly polarized light propagating through an anisotropic saturable absorber is based on the simplest model of the nonlinear absorber \cite{seldon1,seldon2} and does not imply any particular atomic system. This aims to obtain a general pattern of the effects.  

\subsection{Basic equations}

\begin{figure}[t]
\includegraphics[width=.75\columnwidth,clip]{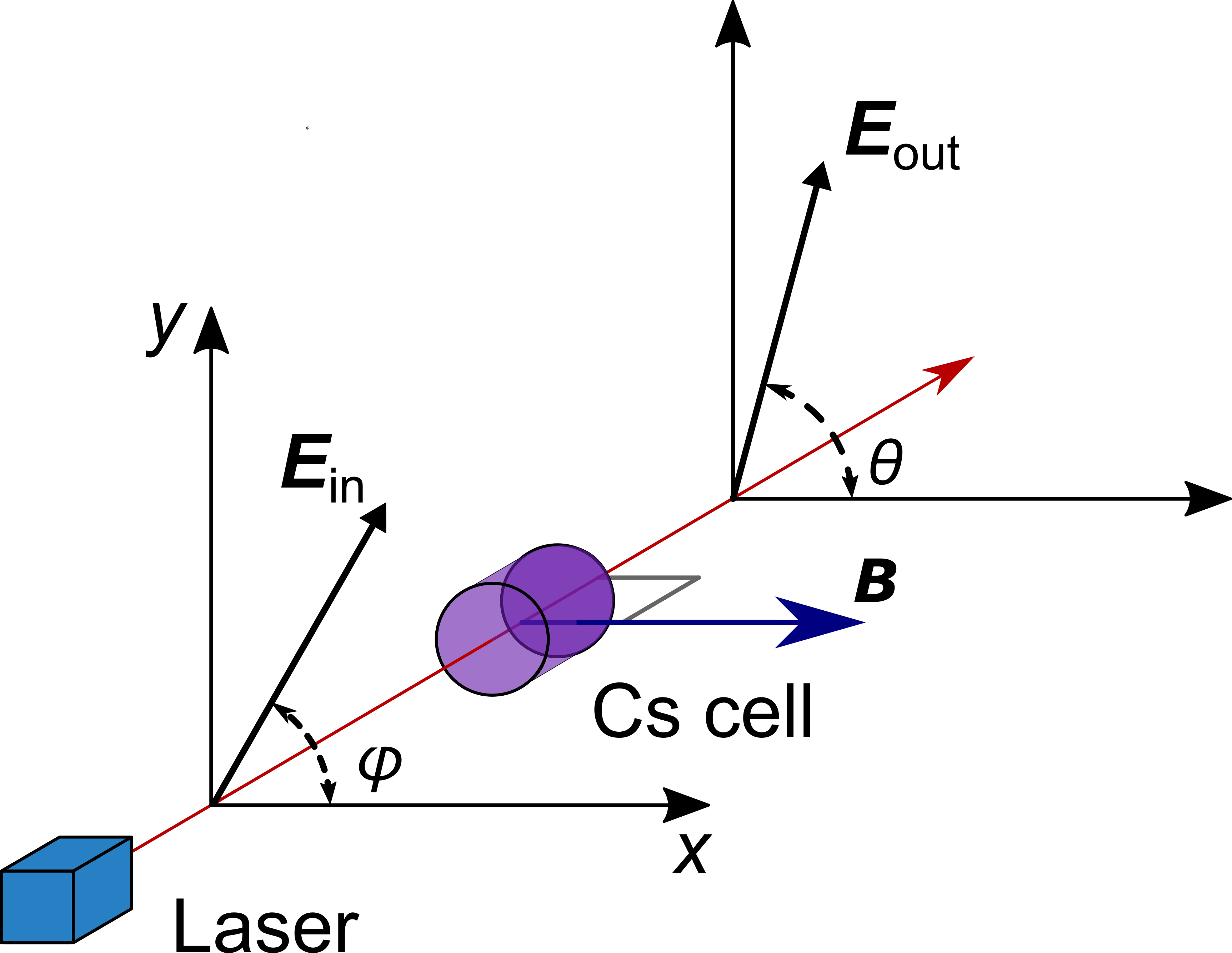}
\caption{Experimental arrangement (the Voigt geometry) and basic notations. External magnetic field $\bm B$ specifies initial anisotropy of the medium, with its axes aligned along $x$ and $y$. Azimuthal angles of the input and output polarization planes ($\varphi$ and $\theta$) are counted from the $x$ axis ($\pi$\ polarization aligned along $\bm B$). 
}\label{fig:1}
\end{figure}

Let the PP of the laser beam incident on the cell with atomic vapor make an angle of $\varphi$ with the transverse magnetic field (see Fig.~\ref{fig:1}). 
In conformity with the aforesaid, the values of absorption in the eigenmodes, as well as the saturation factors describing appropriate effects of bleaching, will be, generally, different.
   
Denote the linear absorption coefficients in the $x$ and $y$ polarizations by $\varkappa_x$ and $\varkappa_y$, with the appropriate optical densities, $D_{x,y} \equiv \varkappa_{x,y} L$, where $L$ is the length of the medium.
Then, the equations describing the intensity variations of the introduced linearly polarized modes (their polarizations are assumed to be aligned along the $x$ and $y$ axes) can be represented, respectively, as follows: 
\begin{equation}
\frac{dI_x}{dz}=-\frac{\varkappa_x I_x}{\sqrt{1+I_x/I^\text{sat}_x}}, \qquad
\frac{dI_y}{dz}=-\frac{\varkappa_y I_y}{\sqrt{1+I_y/I^\text{sat}_y}}.
\label{1}
\end{equation}
Here, $I^\text{sat}_{x}$ and $I^\text{sat}_{y}$ are the saturation intensities in the $x$ and $y$ polarizations, respectively. 
This model assumes isotropic refraction (absence of birefringence).
The initial conditions for these equations are
\begin{equation}
I_x(0)=I_{0}\cos^2\varphi, \qquad I_y(0)=I_{0}\sin^2\varphi,
\label{eq_2}
\end{equation}
where $I_0=I_x+I_y$ is the total light intensity.

In Eqs.\ \eqref{1}, we use, for the law of bleaching of the saturable absorber, the square root dependence $\sqrt{1 + I_{x,y}/I^\text{sat}_{x,y}}$, which implies resonant saturation of the {\it inhomogeneously} (Doppler) broadened transition by a monochromatic laser light~\cite{Letokhov}. 
This circumstance lowers the degree of optical nonlinearity (as compared with the simplest case of the two-level system) and brings certain quantitative changes into the pattern of the response that become substantial only at high levels of saturation. 

The sought-after azimuthal dependences (dependences on the PP azimuth $\varphi$) of the intensities $I_x$ and $I_y$ can be obtained by numerical calculations of Eqs.~\eqref{1}. 
Ultimately, the values of transmission coefficient $T$ and the light PP azimuth $\theta$ measured at the exit of the medium are given by  
\begin{equation}
T = \frac{I_x(L) + I_y(L)}{I_0}, \qquad \theta = \arctan{\sqrt\frac{I_y(L)}{I_x(L)}}.
\end{equation}

\begin{figure}[t]
\includegraphics[width=.95\columnwidth,clip]{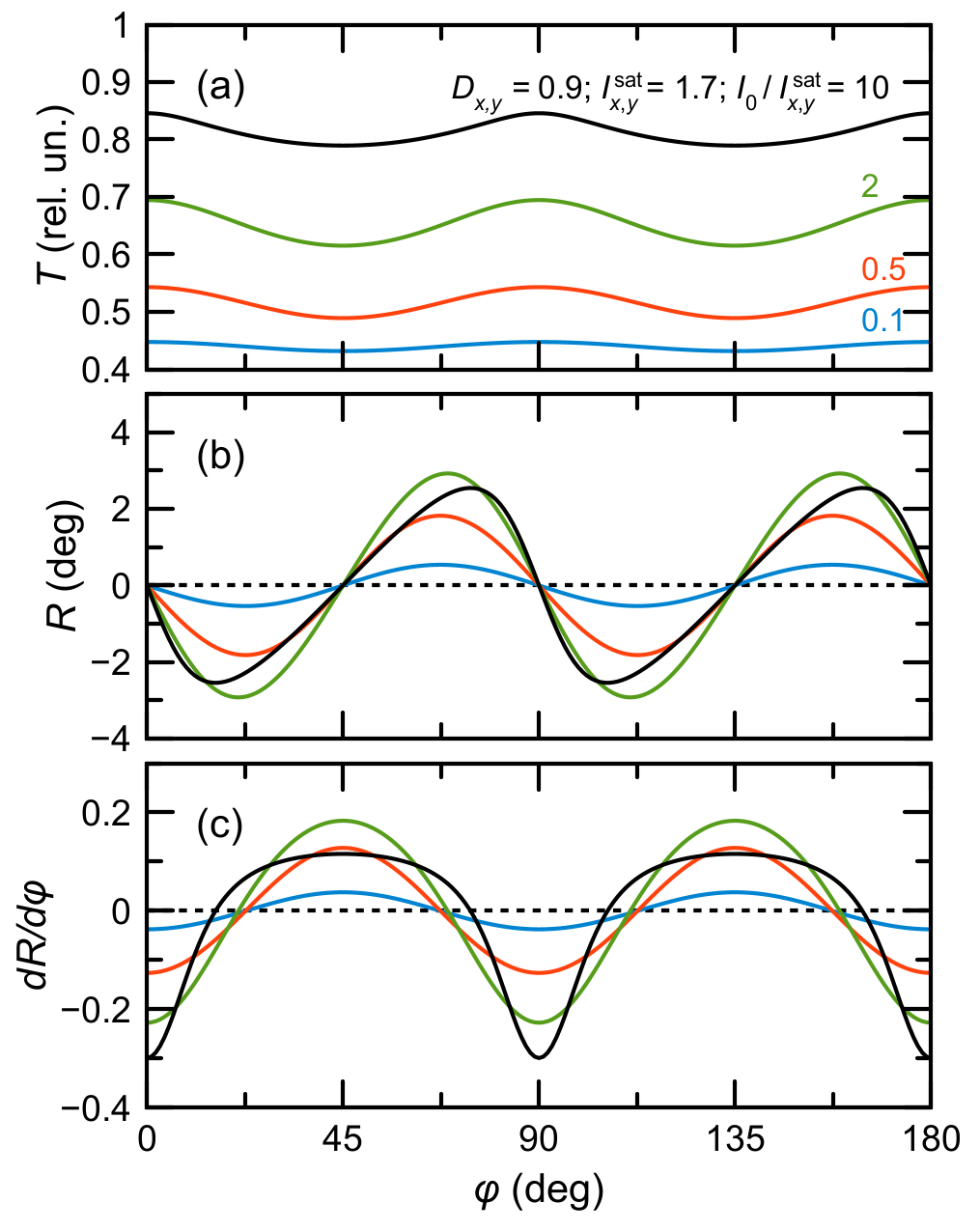}
\caption{\label{fig_2}
Calculated azimuthal dependences of (a) transmission $T(\varphi)$, (b) of the PP rotation angle $R(\varphi)$, and (c) of its derivative $dR/d\varphi$ for a linearly polarized light transmitted through the anisotropic saturable absorber with the identical nonlinearities in the two eigenmodes. 
The parameters of the medium, $D_{x,y} = 0.9$, $I^\text{sat}_{x,y}=1.7$, and 
$I_0/I^\text{sat}_{x,y} = 0.1$ (blue curves), $0.5$ (red curves), $2$ (green curves), and $10$ (black curves) 
were used in Eqs.~\eqref{1} with the initial conditions \eqref{eq_2}. 
}
\end{figure}

\begin{figure}[t]
\includegraphics[width=.95\columnwidth,clip]{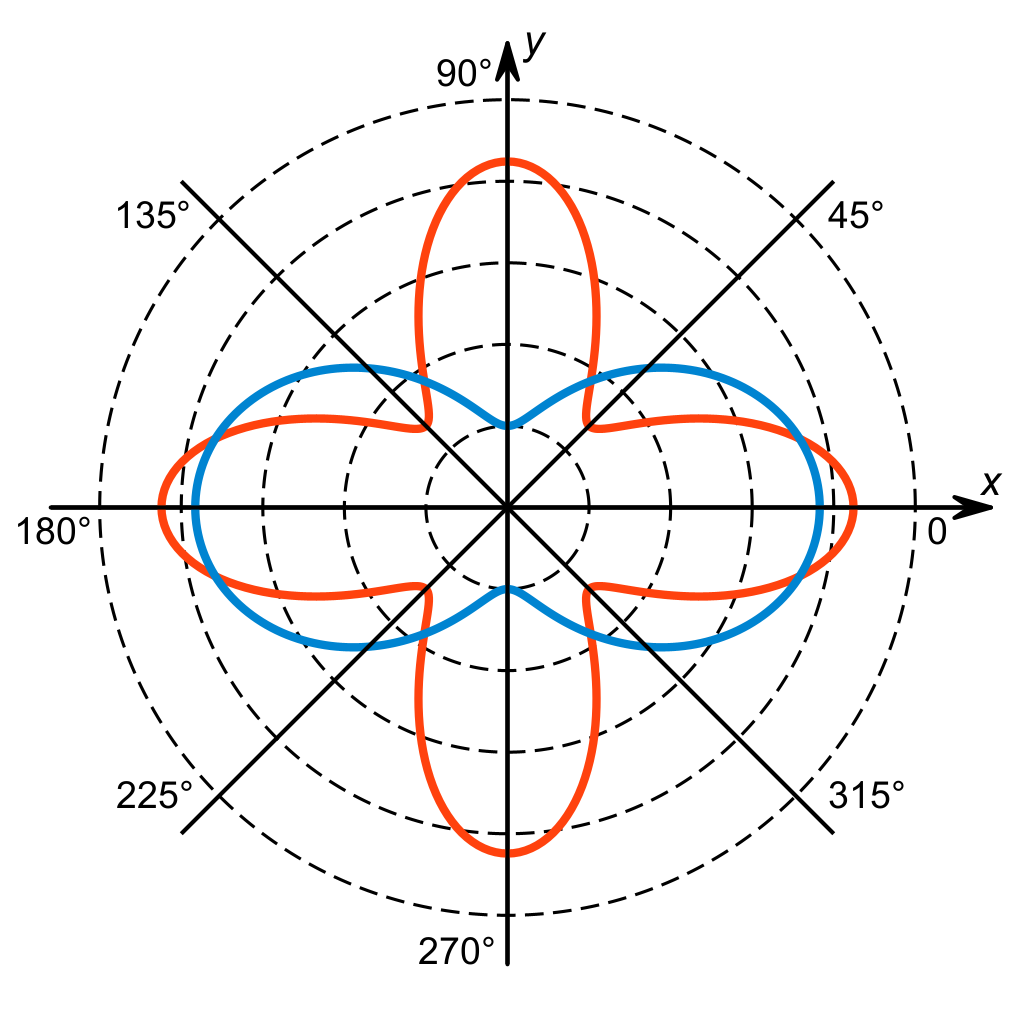}
\caption{\label{fig_3} Schematic representation of angular dependences of transmission for the case of pure dipole (blue curve) and pure quadrupole (red curve) dichroism. The angle $\varphi$ is counted from the magnetic field direction ($x$~axis).
}
\end{figure}

\subsection{Results of calculations}

Figure \ref{fig_2} shows, as an example, the azimuthal dependences of transmission $T(\varphi$) [Fig.\ \ref{fig_2}(a)], the PP rotation, $R(\varphi)$, with $R=\theta-\varphi$ [Fig.\ \ref{fig_2}(b)], and the derivative $d{R}/d\varphi$ [Fig.\ \ref{fig_2}(c)] calculated for certain realistic values of the nonlinearity parameters equal for two eigenaxes ($I^\text{sat}_x$ = $I^\text{sat}_y$ and $D_x = D_y$). 
As has already been mentioned, all these effects are possible only due to the nonlinearity of the medium. 
In linear optics, the equivalence of the two eigenmodes automatically means isotropy of the medium.
%\looseness=1

These dependences demonstrate the most substantial features of the anisotropic saturable absorber under consideration.  
As one can see from Fig.\ \ref{fig_2}(a), the absorption reveals a fairly strong dichroism with the $90^\circ$ periodicity with respect to angle $\varphi$. 
This type of dichroism, which is known to be forbidden for linear optics of anisotropic media, is referred to here as the {\it quadrupole dichroism} (to distinguish it from the conventional {\it dipole dichroism}). 
%\looseness=1

Figure\ \ref{fig_3} illustrates schematically, in polar coordinates, the distinction between the azimuthal dependences of transmission for the case of pure dipole and pure quadrupole dichroism. 
In the general case of the anisotropic saturable absorber, this azimuthal dependence is described by a superposition of the two contributions with the second- and fourth-order symmetry axes.
%\looseness=1

Another magneto-optical effect revealed under these conditions is the PP rotation shown in Fig.~\ref{fig_2}(b). 
In our notation, it is measured as the difference between $\theta$ and $\varphi$. 
Unlike the conventional Faraday effect, this rotation is controlled by the difference in absorption for two linearly polarized normal modes, rather than by the difference in refraction for two modes of circular polarization, and is {\it anisotropic}, i.e., depends on the input azimuth $\varphi$. 
The specific features of this rotation are better seen from the angular dependences of the derivative $dR/d\varphi$ [see Fig.\ \ref{fig_2}(c)], which may be considered as a factor of amplification, $K$, of a small variation of $\varphi$ by the medium, $K = 1 + dR/d\varphi$. 
The fourth-order symmetry of these dependences clearly distinguishes these effects from similar effects in conventional dichroic media with no optical nonlinearity.    
Note that high transparency of the medium in the region of the greatest amplification factors (Fig.\ \ref{fig_2}) favorably distinguishes this quadrupole effect of the PP rotation gain from a similar effect of linear optics \cite{rev,glaz}, which is accompanied by the strong attenuation of the light intensity. Due to this attenuation, the effect of the PP rotation gain in linear optics appears to be useless from the view point  of improving the shot-noise-limited polarimetric sensitivity, and may be helpful only for suppressing excess noise in polarimetric measurements\ \cite{depression}.   
Here, as one can see, the situation is essentially different; the effect of the PP rotation amplification can be observed under conditions of fairly strong bleaching and therefore may have practical significance.
%\looseness=2

\begin{figure}[t]
\includegraphics[width=.95\columnwidth,clip]{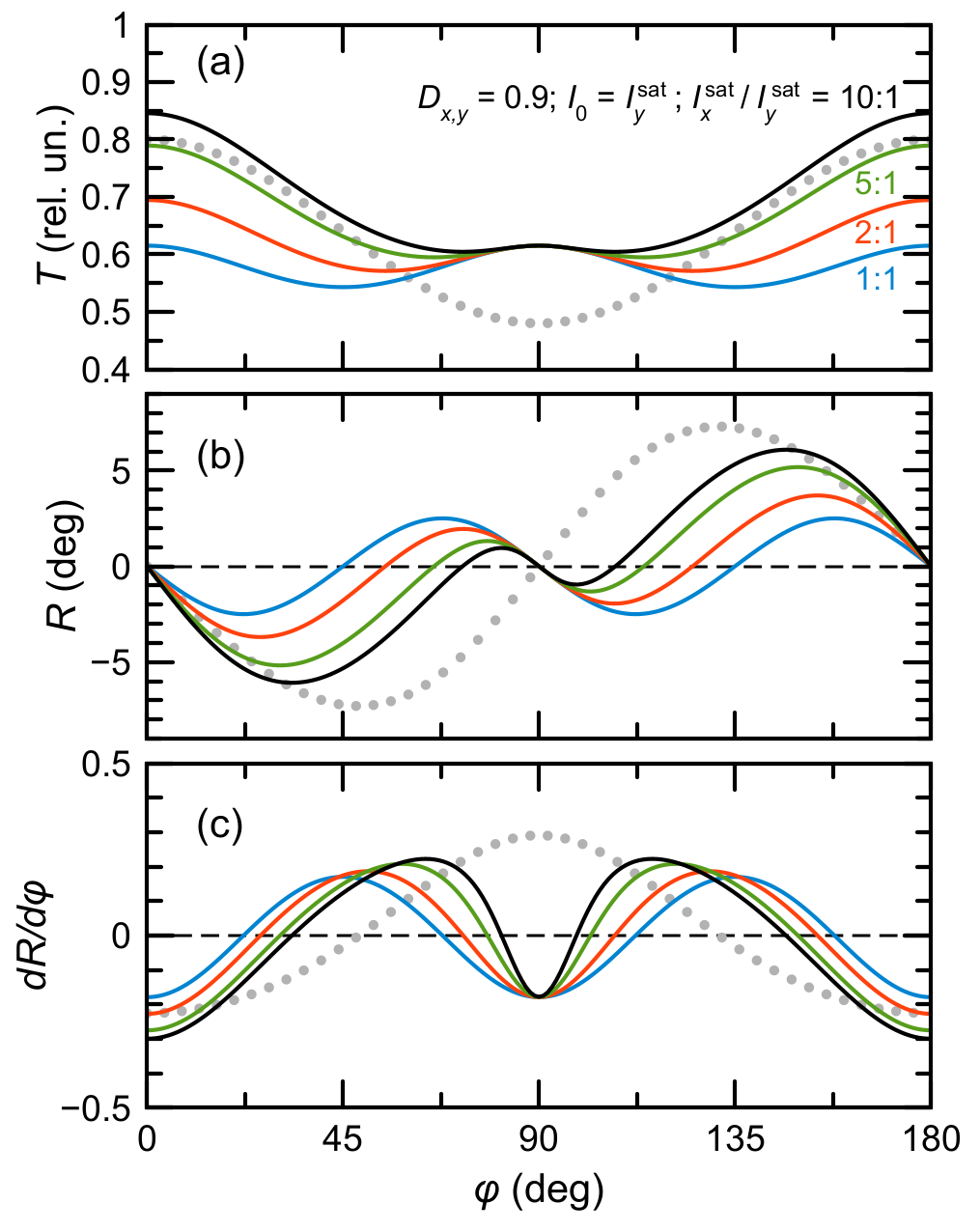}
\caption{Calculated azimuthal dependences of (a) the transmission $T({\varphi})$, (b) {PP} rotation angle $R(\varphi)$, and (c) its derivative $dR/d\varphi$ for a linearly polarized light transmitted through the anisotropic saturable absorber with unequal saturation intensities in the two eigenmodes. The ratios $I^\text{sat}_x / I^\text{sat}_y$ are 1 (blue curves), 2 (red curves), 6 (green curves), and 20 (black curves), and $D_{x,y}=0.9$; $I_0/I^\text{sat}_y = 1$. The gray dotted lines, in the plots, show schematically behavior of these characteristics for a pure dipole dichroism. 
}\label{fig_4}
\end{figure}

The changes in the above azimuthal dependences that occur in more realistic situations, when the nonlinearities in the two eigenmodes $x$ and $y$ are different, are illustrated in Fig.\ \ref{fig_4}. 
The gray dotted curves depict the typical behavior of these characteristics for the case of conventional (dipole) dichroism in a linear medium. 
As was already mentioned, azimuthal dependences of the quadrupole effects dramatically differ from the dichroism-related effects in linear optics: 
in the presence of the quadrupole dichroism, the greatest amplification of the PP rotation occurs in the vicinity of $\varphi = \pm 45^\circ$, rather than at $\varphi = 0$ or $90^\circ$. 
It should be also noticed that, in the presence of quadrupole dichroism, the derivative of the PP rotation angle [Fig.\ \ref{fig_4}(c)] becomes negative for both eigenmodes.  At these points, the medium behaves more like a polarizer and suppresses oscillations of the input-light PP azimuth.
%\looseness=1

One can also notice that, while the dichroism (azimuthal dependence of the transmission) [see Fig.\ \ref{fig_4}(a)] is strongly affected by the introduced asymmetry (due to the admixture of the dipole dichroism), no substantial changes are seen in magnitudes of the PP rotation [see Fig.\ \ref{fig_4}(b)] and their derivatives [see Fig.\ \ref{fig_4}(c)]. The simplest explanation of this curious effect may be given using symmetry consideration: the fourth-order symmetry of the effect  is provided only by the quadrupole dichroism. \looseness=1

\subsection{Additional considerations}

Thus, we see that the only characteristics needed to obtain azimuthal dependences of transmission and PP rotation of such a medium is the law of the absorption saturation.   
Note, however, that if we do not have exact analytical law for the bleaching, the sought-after angular dependences can be found just by using experimentally measured functions of the intensity-dependent transmissions $k_x(I)$ and $k_y(I)$ for the two eigenmodes $x$ and $y$ 
\begin{equation}
I_x = I k_x(I), \qquad
I_y = I k_y(I),
\end{equation}
where $\varphi$ equals $0$ and $\pi/2$ for these two equations, respectively.

Through these functions, angular dependences of the intensity $I(\varphi)$ and PP azimuth $\theta(\varphi)$ of the transmitted light can be represented as 
\begin{equation} 
\begin{split}
I(\varphi) = I_x  + I_y =I_0\left[\cos^2{\varphi}\;  k_x(I_0 \cos^2{\varphi})\right. \\ 
+ \left.\sin^2{\varphi}\; k_y(I_0 \sin^2{\varphi})\right], \qquad 
\theta = \arctan \sqrt \frac{I_y}{I_x}.
\end{split}
\label{eq_5}
\end{equation}
This approach, applicable when the functions  $k_x(I)$ and $k_y(I)$ are known, should give adequate results when our assumption about independent oscillators is valid. 
A discrepancy between the experimental data and the results of the calculations may indicate the incorrectness of the above assumption.

\section{Symmetry considerations}
\label{sec:III}

The above results, obtained in the framework of a simplified model of independent ensembles of saturable absorbers, can be analyzed from the viewpoint of more general symmetry-related considerations. 
The components of the electric field $\bm E$ transform according to the $\mathcal D_1^-$ irreducible representation of the full rotational symmetry group $D$. 
Their combinations of odd power $n$ transform according to the representation $\mathcal D_1^-\oplus{\mathcal D_3^-}\oplus\ldots\oplus\mathcal D_n^-$, so there is only one linearly independent contribution of the given power to the polarization $\bm P$ of the medium, which belongs to $\mathcal D_1^-$ representation. 
It is proportional to $E^{n-1}\bm E$. Therefore, in all powers in $\bm E$, in the absence of a magnetic field, the polarization has the form $\chi_0(E^2)\bm E$.

The components of the magnetic field $\bm B$ belong to $\mathcal D_1^+$ representation. Its first power (and all odd powers) do not contribute to the polarization in this geometry. Its components, to the second power, transform according to the representation {$\mathcal D_0^+\oplus D_2^+$}, so there {can be up to three} contributions to polarization in each odd power of $\bm E$. Then all symmetry-allowed contributions can be combined to the expression
\begin{multline}
    \label{eq:P}
    \bm P=\chi_0(E^2)\bm E+\chi_2(E^2)B^2\bm E\\
    +\chi_d(E^2)(\bm{EB})\bm B+\chi_q(E^2)(\bm{EB})^2\bm E.
\end{multline}
The first line here describes the isotropic nonlinear polarizability, while the two terms in the second line describe dipole and quadrupole dichroism.

Note that the even powers of $\bm E$ do not contribute to the polarization because they do not oscillate at the frequency of the incident field. The contributions to $P_{x,y}$ linear in the light wave vector are also forbidden in this geometry, due to the inversion symmetry, which requires for them even powers of $\bm E$.

Assuming isotropic refraction, the linearly polarized incident light remains linearly polarized during the propagation. Then it can be characterized by the intensities $I_{x,y}$ of the components polarized along and perpendicular to the magnetic field (the $x$ axis). From Eq.~\eqref{eq:P} we obtain general equations for them:
\begin{subequations}
  \label{eq:dI}
  \begin{equation}
    \frac{\d I_{x}}{\d z}=-\alpha_0(I)I_x-\alpha_d(I)I_x-\alpha_q(I)I_x^2,
  \end{equation}
  \begin{equation}
    \frac{\d I_{y}}{\d z}=-\alpha_0(I)I_y,
  \end{equation}
\end{subequations}
where $I=I_x+I_y$ is the total intensity, $\alpha_0(I)$ describes isotropic part of the absorption, and $\alpha_{d,q}(I)\propto B^2$ describe dipole and quadrupole dichroism in the absorption.

Neglecting the latter ones, the isotropic part of the intensity $I_0(z)$ satisfies the equation
\begin{equation}
  \frac{\d I_0}{\d z}=-\alpha_0(I_0)I_0.
\end{equation}
Now, the two intensities can be written as $I_x=I_0\cos^2\phi+\delta I_x$, $I_y=I_0\sin^2\phi+\delta I_y$. In the first order in $\alpha_d$ and $\alpha_q$ we obtain from Eqs.~\eqref{eq:dI},
\begin{subequations}
  \label{eq:delta_I}
  \begin{multline}
    \frac{\d\delta I_x}{\d z}=-\alpha_0'(I_0)I_0(\delta I_x+\delta I_y)\cos^2\phi\\
    -\alpha_0(I_0)\delta I_x-\alpha_d(I_0)I_0\cos^2\phi-\alpha_q(I_0)I_0^2\cos^4\phi,
  \end{multline}
  \begin{equation}
    \frac{\d\delta I_y}{\d z}=-\alpha_0'(I_0)I_0(\delta I_x+\delta I_y)\sin^2\phi-\alpha_0(I_0)\delta I_y,
  \end{equation}
\end{subequations}
where the prime denotes the derivative over $I_0$. Summing up these equations we find for $\delta I=\delta I_x+\delta I_y$
\begin{multline}
  \frac{\d\delta I}{\d z}=-\alpha_0'(I_0)I_0\delta I-\alpha_0(I_0)\delta I\\-\alpha_d(I_0)I_0\cos^2\phi-\alpha_q(I_0)I_0^2\cos^4\phi,
\end{multline}
with the {solution
\begin{multline}
  \label{eq:delta_I_tot}
  \delta I(z)=-\int\limits_0^z\left[\alpha_d(z')I_0(z')\cos^2\phi+\alpha_q(z')I_0^2(z')\cos^4\phi\right]\\\times\exp\left\{-\int\limits_{z'}^z\left[\alpha_0'(z'')I_0(z'')+\alpha_0(z'')\right]\d z''\right\}\d z',
\end{multline}
where $\alpha_{0,d,q}(z)=\alpha_{0,d,q}\left[I_0(z)\right]$.}

Now we substitute this expression in Eqs.~\eqref{eq:delta_I} and obtain from it the final solution
\begin{subequations}
  \begin{multline}
    \label{eq:Ix1}
    \delta I_x(L)=-\int\limits_0^L\left[\alpha_0'(z)I_0(z)\delta I(z)\cos^2\phi
    \right.\\\left.
    +\alpha_d(z)I_0(z)\cos^2\phi+\alpha_q(z)I_0^2(z)\cos^4\phi\right]
    \\\times
    \exp\left(-\int\limits_{z}^L\alpha_0(z')\d z'\right)\d z,
  \end{multline}
  \begin{multline}
    \label{eq:Iy1}
    \delta I_y(L)=-\int\limits_0^L\alpha_0'(z)I_0(z)\delta I(z)\sin^2\phi
    \\\times
    \exp\left[-\int\limits_{z}^L\alpha_0(z')\d z'\right]\d z.
  \end{multline}
\end{subequations}
The angular dependence of these quantities has the form
\begin{subequations}
  \label{eq:delta_I_xy}
  \begin{equation}
    \delta I_x=(I_d\cos^2\phi+I_q\cos^4\phi)\cos^2\phi+I_d'\cos^2\phi+I_q'\cos^4\phi,
  \end{equation}
  \begin{equation}
    \delta I_y=(I_d\cos^2\phi+I_q\cos^4\phi)\sin^2\phi,
  \end{equation}
\end{subequations}
where the contributions $I_d$, $I_d'$ and $I_q$, $I_q'$ are proportional to $\alpha_d$ and $\alpha_q$, respectively.

The transmission coefficient is given by $T=[I_0(L)+\delta I(L)]/I_0(0)$ and has the form
\begin{equation}
  \label{eq:delta_T2}
  T=T_0+(T_d+T_d')\cos^2\phi+(T_q+T_q')\cos^4\phi,
\end{equation}
where $T_0=I_0(L)/I_0(0)$, $T_{d,q}=I_{d,q}(L)/I_0(0)$, and $T_{d,q}'=I_{d,q}'(L)/I_0(0)$.

The rotation angle of the PP is determined by $R=[\delta I_y(L)\cos^2\phi-\delta I_x(L)\sin^2\phi]/[I_0(L)\sin2\phi]$ and equals
\begin{equation}
  \label{eq:delta_phi2}
  R=-\frac{T_q'+2T_d'}{8T_0}\sin2\phi-\frac{T_q'}{16T_0}\sin4\phi.
\end{equation}
Of note, this expression directly yields the amplification factor of the polarization plane rotation $K$ introduced above. It depends, in particular, on the transmission coefficients $T_d'$ and $T_q'$, which are related to the dipole and quadrupole dichroism, respectively. The latter can both enhance and suppress the amplification factor. In the experimentally important case of $-4<T_q'/T_d'<-2$ the gain in the amplification factor due to the quadrupole dichroism is $G=(1-T_q'/T_d')/3$ and it increases with  increasing  $|T_q'/T_d'|$.

From this general analysis, one can draw a few conclusions: (i) Quadrupole anisotropy requires at least the third power of the electric field [see Eq.~\eqref{eq:P}] and is forbidden in linear optics. (ii) In the second order in the magnetic field and for weak anisotropy, there are only second and fourth angular harmonics in the transmission coefficient (dipole and quadrupole dichroism, respectively) and in the PP rotation. The higher harmonics, present in the previous section, require higher powers of the magnetic field. (iii) There are three contributions to the angular dependence of the transmission coefficient and two contributions to the PP rotation, which are all linearly independent.

The next section demonstrates  good agreement of the experimental results with these conclusions.

\section{Experimental illustration}
\label{sec:IV}

\begin{figure}
\includegraphics[width=\columnwidth,clip]{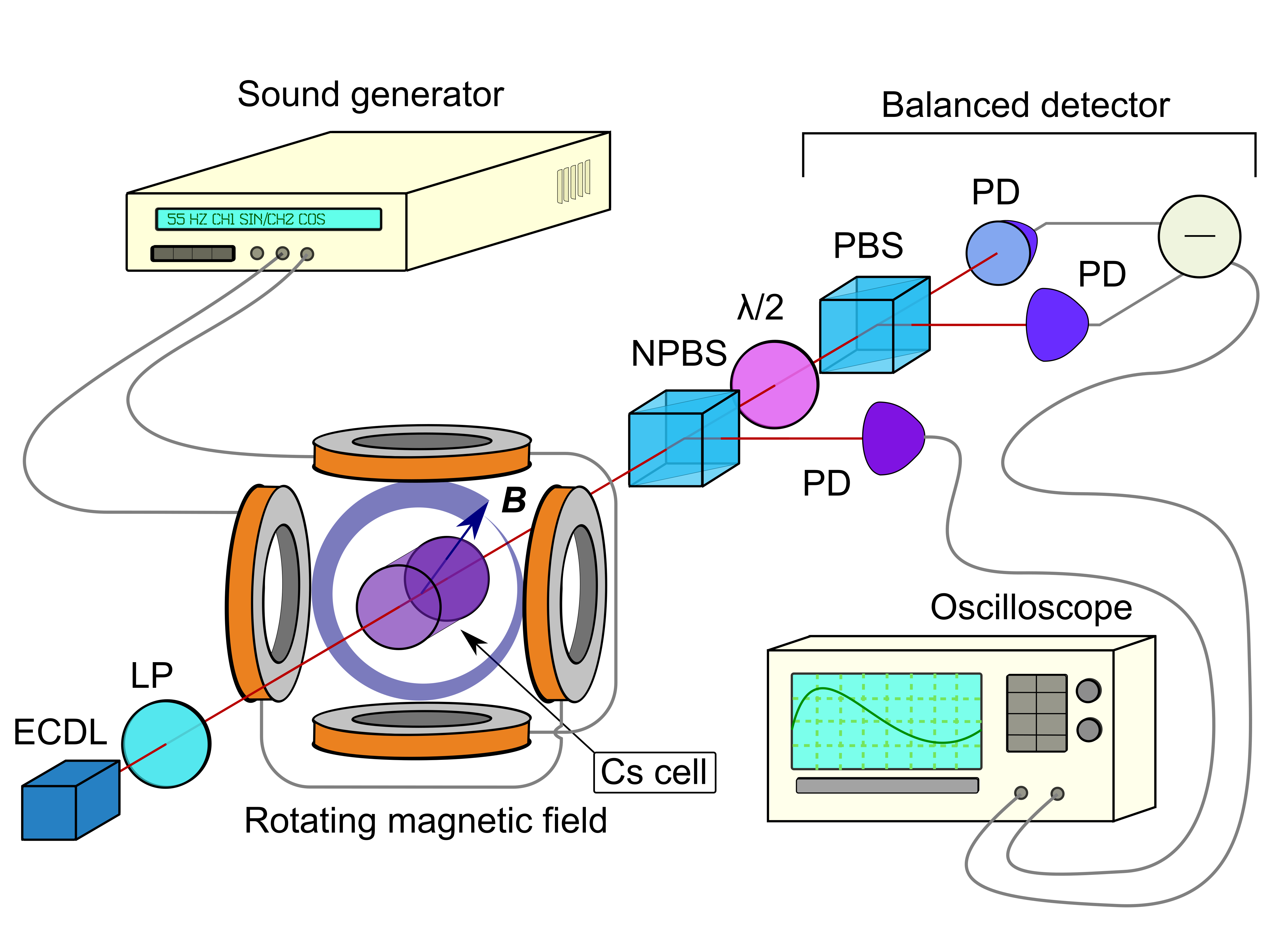}
\caption{Schematic of the experimental setup. ECDL is the
external-cavity diode laser, LP is the linear polarizer, $\lambda/2$ is the half-wave plate, NPBS and PBS are the nonpolarizing and polarizing beamsplitters, respectively, and PDs are the photodetectors.
}\label{fig_schematic}	
\end{figure}

In this section, we present an experimental illustration of the described effects of quadrupole dichroism and anisotropic PP rotation. 
The measurements are performed on cesium vapor in a cell without any buffer gas. 
The magnitude and general pattern of these effects is evidently controlled by specific parameters of this particular system and therefore can be used to extract their values. 
However, we do not consider these effects here as a tool of research and restrict ourselves only to the demonstration of the correctness of the predicted phenomenology of the effects and validity of our simple theoretical model. 
A more accurate quantitative analysis of these effects, as applied to a specific atomic system, will be presented elsewhere.
\looseness=-1

The schematic of the experimental setup is shown in Fig.~\ref{fig_schematic}.  
As a light source, we use the external-cavity diode laser (Toptica DLC DL Pro) tunable within $\sim$$20$\ GHz at the wavelength $\lambda = 852.35$\ nm. 
The laser beam, $2$~mm in diameter, tuned in resonance with one of the hyperfine components of the $D2$ line of Cs, passes through the cell with saturated Cs vapor, $20 \times 20$\ mm in size. 
The temperature of the vapor cell is maintained at $62^\circ$~C. 
Measurements of the azimuthal dependences of the transmission and PP rotation in a transverse magnetic field are performed by rotating the external magnetic field, rather than the light polarization, around the light propagation direction. 
For this purpose, we use two pairs of coils aligned in orthogonal directions and fed at a frequency of $55$ Hz from two quadrature outputs of a sound generator.  
In this case, the azimuth of the input light polarization is fixed and the rotation angle is directly measured by the imbalance signal of the balanced detector normalized by the intensity of the transmitted light. 

\begin{figure}[t]
\includegraphics[width=.95\columnwidth,clip]{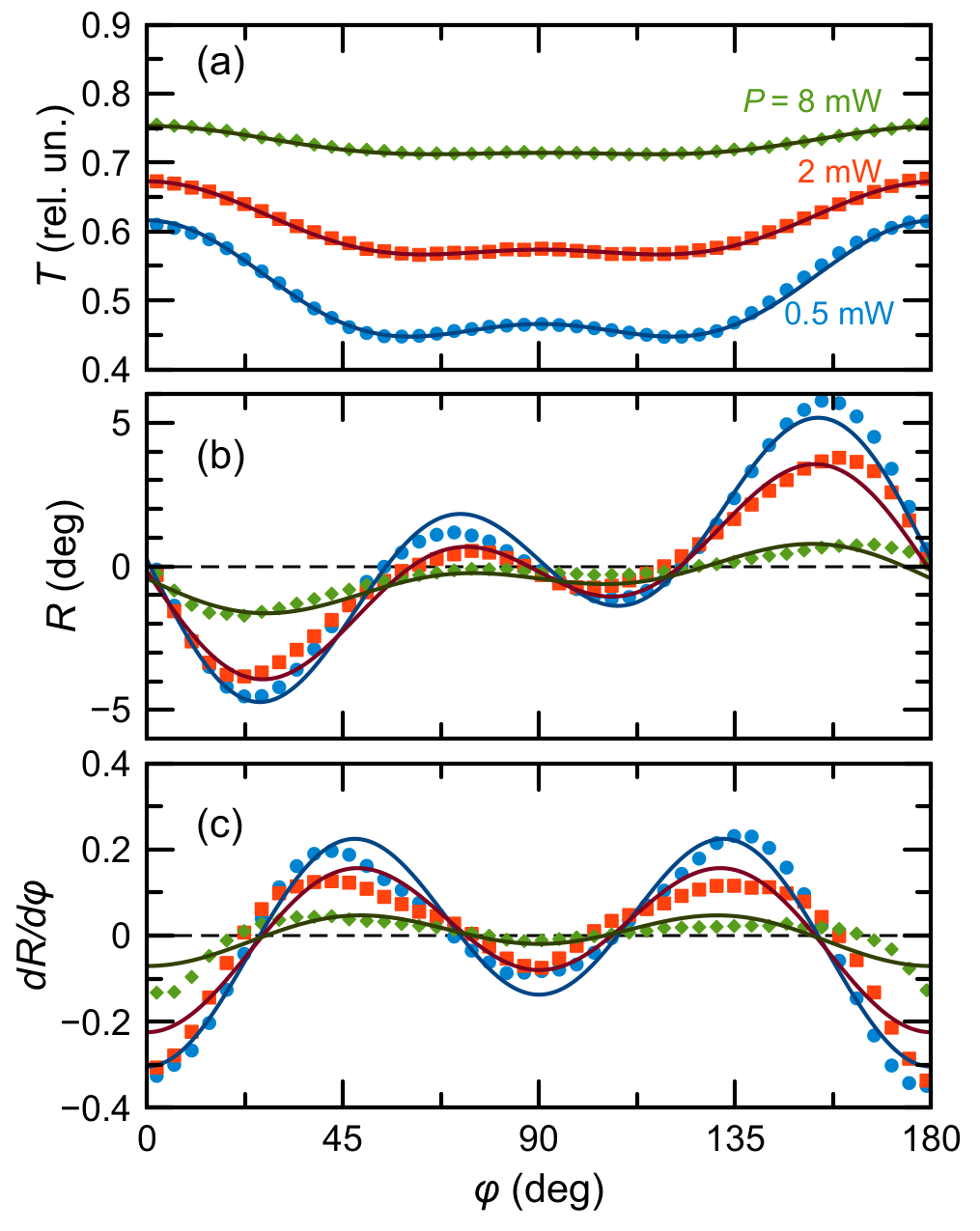}
\caption{Experimental azimuthal dependences of (a) the transmission $T(\varphi)$, (b) the PP rotation $R(\varphi)$, and (c) its derivative $dR/d\varphi$ measured in cesium vapor for different light power (markers). Solid lines are the fits after Eqs.\ \eqref{eq:delta_T2} and \eqref{eq:delta_phi2} with the parameters listed in Table\ \ref{tab_params}.}
\label{fig_6_experimental}	
\end{figure}

\begin{table}[b]
\caption{The list of parameters used for the fits in Fig.\ \ref{fig_6_experimental}.}
\begin{tabular}{c|c|c|c|c|c|c|c}
$P$ & \multirow{2}{*}{$T_0$}    & \multirow{2}{*}{$T_{q}'$}   & \multirow{2}{*}{$T_{d}'$}   & 
\multirow{2}{*}{$T_q$}     & \multirow{2}{*}{$T_d$}     & \multirow{2}{*}{$\left|\frac{T_q'}{T_d'}\right|$} & \multirow{2}{*}{$G$} \\
(mW) & & & & & & & \\
\hline
\hline
$0.5$    & $0.4658$ & $0.4108$   & $-0.1277$  & $-0.1128$ & $-0.0201$ & $3.2168$      & $1.4$ \\
$2.0$    & $0.5736$ & $0.3492$   & $-0.0919$  & $-0.1818$ & $\phantom{-}0.0236$  & $3.7993$      & $1.6$ \\
$8.0$    & $0.7140$ & $0.1285$   & $-0.0276$  & $-0.0677$ & $\phantom{-}0.0057$  & $4.6521$      & $1.9$ 
\end{tabular}
\label{tab_params}
\end{table}

Figure\ \ref{fig_6_experimental} shows the experimentally measured azimuthal dependences of transmission, PP rotation, and derivative of the PP rotation for the laser light tuned in resonance with the long-wavelength ground-state hyperfine component ($F = 3$) of the $D2$ line of Cs.  
In these experiments, the factors affecting contrast of the dependences (linear optical density and light power) are not strong enough, as assumed in the above calculations, but, still, the most important features of the quadrupole anisotropy are clearly seen. 
Specifically, the dichroism-related PP rotation turns into zero not only at $\varphi = 0$ and $90^\circ$ like in the media with conventional dipole dichroism, but also at the points close to $\pm 45^\circ$.  

\begin{figure}[t]
\includegraphics[width=.95\columnwidth,clip]{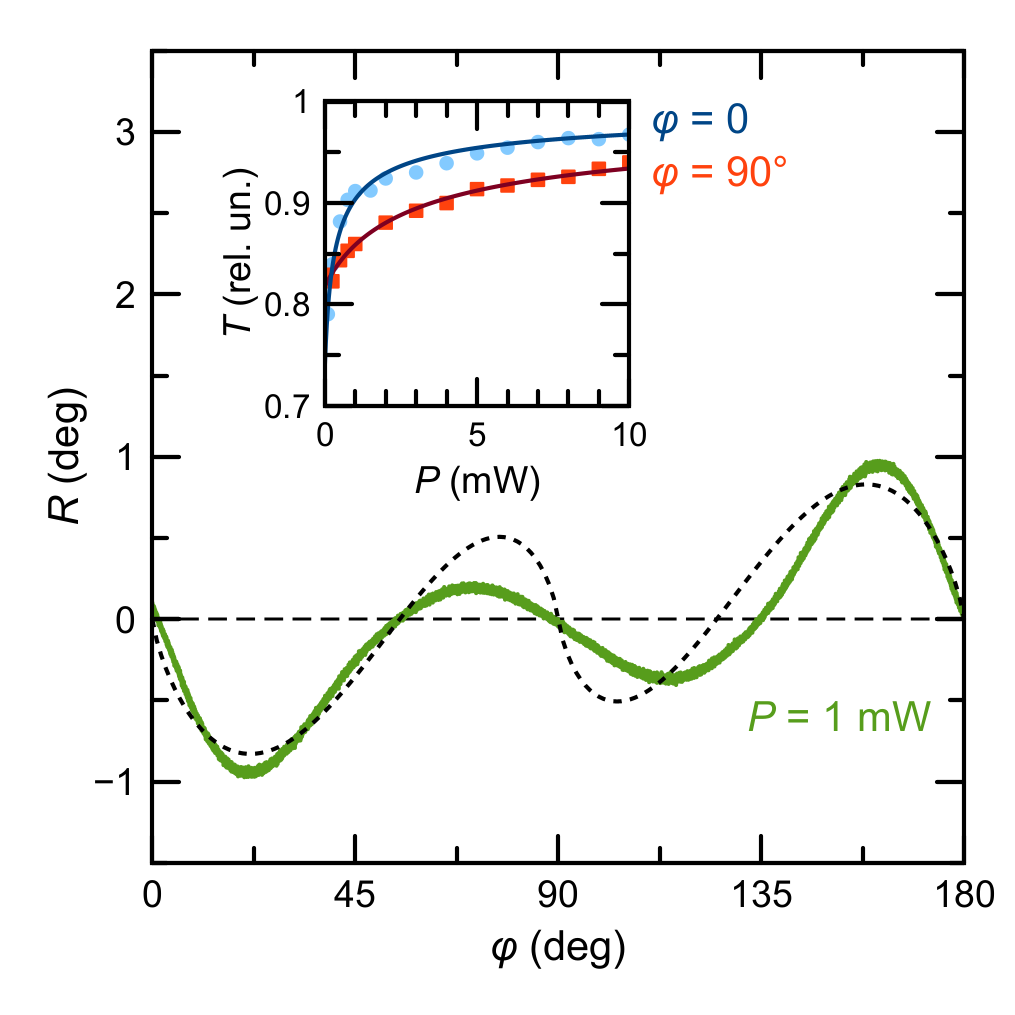}
\caption{Comparison of the azimuthal dependence of the PP rotation, $R(\varphi)$, measured for $P = 1$\ mW (green curve) with that obtained using Eq. (5) (black dotted line). 
The measured intensity-dependent transmission of the vapor at $\varphi = 0$ (blue circles) and $\varphi = 90^\circ$ (red squares) is shown in the inset. 
}
\label{fig_7_experimental}	
\end{figure}

It is important that, while the derivatives $dR/d\varphi$, at $\varphi = 0$ and $90^\circ$, are negative, and small rotations of the PP at the entrance of the cell appear to be suppressed at the exit, at the points close to $\pm 45^\circ$, the derivatives $dR/d\varphi$ are seen to be essentially positive, and the quadrupole dichroism, under these conditions, can be used for the amplification of a small PP rotation. One can nicely fit the curves in Fig.~\ref{fig_6_experimental} with Eqs.~\eqref{eq:delta_T2} and~\eqref{eq:delta_phi2}, which gives the parameters listed in Table~\ref{tab_params}. From these parameters, we calculate the gain in the PP rotation amplification given in the last column in Table~\ref{tab_params}. One can see that the quadrupole dichroism plays an important role in the amplification of the PP rotation in our system with the gain as strong as almost two times.

As mentioned above, the polarization transformation of the light transmitted by such a saturable absorber is completely controlled by the laws of bleaching of the medium along its eigenmodes.  
In Fig.\ \ref{fig_7_experimental}, we present, as an example, the comparison of an experimental azimuthal dependence with that calculated by using Eq.\ \eqref{eq_5} with the experimentally measured intensity-dependent transmissions along the two eigenmodes (see inset in Fig.\ \ref{fig_7_experimental}). An insignificant distinction between these curves, as we believe, is related to the dynamics of the ground-state hyperfine pumping ignored in our model.  

Note that, despite the simplicity of the model presented in Sec.\ \ref{sec:II}, our experimental data well correlate, both qualitatively and quantitatively, with the results of calculations. In particular, all experimental curves show an inflection at $\theta = 90^\circ$, thus indicating a contribution of the quadrupole dichroism.  For the parameters chosen in Fig.\ \ref{fig_4}(b), the experimental curve of Fig.\ \ref{fig_7_experimental} best matches the calculations at $I_x^\text{sat}/ I_y^\text{sat} = 2$. We think it does not make sense to accurately fit experimental data to our simplified calculations. 

\section{Discussion}
\label{sec:V}

As known from the linear optics, an anisotropic layer cannot have two or more physically equivalent directions. 
If it has, it is isotropic. 
This is not the case in nonlinear optics, when the optical anisotropy of the medium can reveal symmetry of higher order. 
The quadrupole effects of the dichroism considered in this paper using the example of cesium vapor belong to this type of phenomena and can be observed in a broad class of optical media. 
The essence of the matter can be explained by using the tensor approach. 
In linear optics, the susceptibility of any anisotropic medium is described by the second-rank tensor. 
For equal eigenvalues, this tensor turns into a scalar, and the medium becomes isotropic. 
The optical susceptibility of a {\it nonlinear} medium is described by a tensor of higher rank having a larger number of invariants and a characteristic surface of higher order. 
Without entering into the details of this symmetry-related problem, we can say that the higher rank of the susceptibility tensor of the medium, the higher angular harmonics of the anisotropy can be detected. 
%\looseness=-1

The simple model of optical nonlinearity presented in Sec.~\ref{sec:II} captures the main observable effects related to the quadrupole dichroism. At the same time, it assumes an ensemble of anisotropic absorbers, which may be relevant for atomic vapors affected by a strong magnetic field only or for crystals without full rotational symmetry. Thus, the symmetry analysis of Sec.~\ref{sec:III} complements this model by the derivation of the expressions for the anisotropy of the transmission and PP rotation taking the magnetic field into account to the second order only. We note that accounting for the higher powers of the magnetic field yields the higher harmonics in Eqs.~\eqref{eq:delta_T2} and~\eqref{eq:delta_phi2}, which reconciles the two approaches.

In the context of symmetry consideration, it makes sense to note that a similar effect of quadrupole dichroism can be observed in the static electric (rather than magnetic) field. It follows from the fact that the even-parity contributions to optical polarization are transformed, in these two cases, identically.

In this work, we chose atomic vapors as a convenient model object for observation of the magnetic quadrupole effect. The nonlinearity of this system is determined by its anisotropic bleaching under conditions of optical pumping. In contrast to the other version of the optical pumping (optical orientation\ \cite{happer,skrot,optor}), optical alignment implies destruction of the axial symmetry of the medium and does not link the effect of the PP rotation with the magnetic gyrotropy of the medium. 
One should note that the described effect of the quadrupole dichroism is not unique and can be observed in other saturable absorbers whose initial isotropy is perturbed by an external field. 
Examples are dye solutions or glasses with paramagnetic impurities (e.g., with rare-earth or transition-metal ions). 
Of course, from the viewpoint of the magnitude of the effect and the possibility of its application, each system deserves special consideration.

There are several experimental investigations with close ideology where the angular polarization dependence of nonlinear effects was used to reveal the hidden anisotropy of cubic crystals. 
In Ref.\ \cite{cr4}, the nonlinear absorption spectroscopy was applied to study the hidden anisotropy of the Cr$^{4+}$:YAG crystal. 
A number of examples of a polarization dependence of the nonlinear refractive index that allow one to reveal the hidden anisotropy of the crystal were given in Ref.\ \cite{zscan}. 
Among these kinds of investigations, the luminescence-based method of detecting hidden anisotropy of cubic crystals with impurity centers can be mentioned~\cite{feofil}. 
In those experiments, the nonlinearity was, in fact, the result of incoherence of the response. 
\looseness=-1

The specific characteristics of the medium considered in this paper are related to the fact that bleaching of the saturable absorber occurs due to the effect of hole burning in the inhomogeneously broadened (Doppler-broadened) line. 
As a result, the nonlinearities of the medium and their manifestations appear to be weaker than for the case of a simple homogeneously broadened absorber. 
However, the light-induced anisotropy of the medium, under these conditions, is reduced to dichroism because birefringence, in the center of the hole, vanishes.  
The latter fact was confirmed by our measurements.

\section{Conclusions}

We studied theoretically and experimentally studied the polarization characteristics of the anisotropic saturable absorber at different parameters of the nonlinearity for its two eigenmodes. 
As an example of such a medium, we chose cesium vapor in a transverse magnetic field resonantly probed at the wavelength of the $D2$ line. 
We showed that, in small magnetic fields, when, in the framework of linear optics, practically no field-induced anisotropy can be detected, the medium becomes strongly anisotropic under conditions of saturation. 
The most spectacular polarization effects observed under these conditions in a linearly polarized light are the specific azimuthal dependences of transmission and PP rotation with $90^\circ$ periodicity. 
This type of dichroism, which we refer to here as quadrupole, is fairly large in magnitude and strongly differs from the known quadratic magneto-optical effects.
%\looseness=-1

Regarding its possible magnetometric applications, this effect is not supposed to be a competitor of conventional methods of the optical magnetometry (see, e.g., Refs.~\cite{budker3, budker2, jackson, makarova}), which are aimed at measuring the magnetic-field magnitude (either through the magnetic resonance frequency or through the Faraday rotation angle). 
The effect of the quadrupole dichroism is evidently more sensitive to the magnetic-field {\it direction} and should be used most efficiently in the compass-type instruments when the field magnitude is not of primary importance.  
%\looseness=+1

The quantitative characteristics of the effect may serve as a source of information about properties of a particular system. A more practical conclusion that can be inferred from the above consideration is that, for an isotropic medium (either in any static perturbating field or not), the presence of a higher-order harmonics ($>$$2$) in azimuthal dependence of its optical response may serve as an indicator of optical nonlinearity.

\acknowledgments

A.A.F., G.G.K, M.Y.P., and V.S.Z. acknowledge Saint-Petersburg State University for the research Project No. 1024022800259-7. 
D.S.S. acknowledges the Foundation for the Advancement of Theoretical Physics and Mathematics ``BASIS.'' 
M.V.P. acknowledges Ioffe Institute for the research Project No. FFUG-2024-0005. The authors are thankful to A.~K.\ Vershovskii for fruitful discussions. M.Y.P. thanks Magicplot Systems, LLC, which provided the software in which the experimental data were processed.
\looseness=-1

%\newpage


\begin{thebibliography}{35}
\bibitem{happer} W.\ Happer, Optical pumping, \href{https://doi.org/10.1103/RevModPhys.44.169}{\rmp\ \textbf{44}, 169 (1972)}.
\bibitem{budker1} D.\ Budker, W.\ Gawlik, D.\ F.\ Kimball, S.\ M.\ Rochester, V.\ V.\ Yashchuk, and A.\ Weis, Resonant nonlinear magneto-optical effects in atoms, \href{https://doi.org/10.1103/RevModPhys.74.1153}{\rmp\ \textbf{74}, 1153 (2002)}. 
\bibitem{nonlinear1} B.\ Ai and  R.\ J.\ Knize, Optical information processing using alkali-metal vapors, \href{https://doi.org/10.1002/adma.19950070317}{Adv.\ Mater.\ \textbf{7}, 319 (1995)}.
\bibitem{nonlinear2} Q.\ Glorieux, T.\ Aladjidi, P.\ D.\ Lett, and R.\ Kaiser, Hot atomic vapors for nonlinear and quantum optics, \href{https://doi.org/10.1088/1367-2630/acce5a}{New\ J.\ Phys.\ \textbf{25}, 051201 (2023)}. 
\bibitem{nonlinear3} P.\ Siddons, Light propagation through atomic vapours, \href{https://doi.org/10.1088/0953-4075/47/9/093001}{J.\ Phys.\ B:\ At.\ Mol.\ Opt.\ Phys.\ \textbf{47}, 093001 (2014)}. 
\bibitem{skrot} G.\ V.\ Skrotskii and T.\ G.\ Izyumova, Optical orientation of atoms and its applications, \href{https://doi.org/10.1070/PU1961v004n02ABEH003331}{Sov.\ Phys.\ Usp.\ \textbf{4}, 177 (1961)}.
\bibitem{anderson} N. Andersen, {\it Orientation and Alignment in Atomic and Molecular Collisions}, Chap.\ 48 in Springer Handbook of Atomic, Molecular, and Optical Physics, 2nd Edition, edited by G. Drake (Springer, Cham, Switzerland, 2023). %ISBN:\ \href{https://isbnsearch.org/isbn/9783030738921}{9783030738921}
\bibitem{EIT1} O.\ S.\ Mishina, M.\ Scherman, P.\ Lombardi, J.\ Ortalo, D.\ Felinto, A.\ S.\ Sheremet, D.\ Kupriyanov, J.\ Laurat, and E.\ Giacobino, Enhancement of electromagnetically induced transparency in room temperature alkali metal vapor, \href{https://doi.org/10.1134/S0030400X1111021X}{Opt.\ Spectrosc.\ \textbf{111}, 583 (2011)}.
\bibitem{EIT3} M.\ Fleischhauer, A.\ Imamoglu, and J.\ P.\ Marangos, Electromagnetically induced transparency: Optics in coherent media, \href{https://doi.org/10.1103/RevModPhys.77.633}{\rmp\ \textbf{77}, 633 (2005)}.
\bibitem{EIT2} R.\ Finkelstein, S.\ Bali, O.\ Firstenberg, and I.\ Novikova, A practical guide to electromagnetically induced transparency in atomic vapor, \href{https://doi.org/10.1088/1367-2630/acbc40}{New\ J.\ Phys.\ \textbf{25}, 035001 (2023)}.
\bibitem{CPT1} J.\ Vanier, A.\ Godone, and F.\ Levi, Coherent population trapping in cesium: Dark lines and coherent microwave emission, \href{https://doi.org/10.1103/PhysRevA.58.2345}{\pra\ \textbf{58}, 2345 (1998)}.
\bibitem{CPT2} H.\ Kim, H.\ S.\ Han, T.\ H.\ Yoon, and D.\ Cho, Coherent population trapping in a configuration coupled by magnetic dipole interactions, \href{https://doi.org/10.1103/PhysRevA.89.032507}{\pra\ \textbf{89}, 032507 (2014)}.
\bibitem{Bloembergen} N.\ Bloembergen, Nonlinear optics and spectroscopy, \href{https://doi.org/10.1103/PhysRevA.54.685}{\rmp\ \textbf{54}, 685 (1982)}.
\bibitem{Boyd} R.\ W.\ Boyd, \textit{Nonlinear Optics}, 4th Edition (Academic Press, San Diego, CA, 2020). %ISBN:\ \href{https://isbnsearch.org/isbn/9780128110027}{9780128110027}
\bibitem{4wm1} T.\ Yajima and Y.\ Taira, Spatial optical parametric coupling of picosecond light pulses and transverse relaxation effect in resonant media, \href{https://doi.org/10.1143/JPSJ.47.1620}{J.\ Phys.\ Soc.\ Jpn.\ \textbf{47}, 1620 (1979)}.
\bibitem{4wm2} K.\ Leo, M.\ Wegener, J.\ Shah, D.\ S.\ Chemla, E.\ O.\ G\"obel, T.\ C.\ Damen, S.\ Schmitt-Rink, and W.\ Sch\"afer, Effects of coherent polarization interactions on time-resolved degenerate four-wave mixing, \href{https://doi.org/10.1103/PhysRevLett.65.1340}{\prl\ \textbf{65}, 1340 (1990)}.
\bibitem{4wm3} A.\ V.\ Trifonov, I.\ A.\ Yugova, A.\ N.\ Kosarev, Ya.\ A.\ Babenko, A.\ Ludwig, A.\ D.\ Wieck, D.\ R.\ Yakovlev, M.\ Bayer, and I.\ A.\ Akimov, Long-lived photon echo induced by nuclear spin fluctuations in charged InGaAs quantum dots, \href{https://doi.org/10.1103/PhysRevB.109.L041406}{\prb\ \textbf{109}, L041406 (2024)}.
\bibitem{2h1} P.\ A.\ Franken, A.\ E.\ Hill, C.\ W.\ Peters, and G.\ Weinreich, Generation of optical harmonics, \href{https://doi.org/10.1103/PhysRevLett.7.118}{\prl\ \textbf{7}, 118 (1961)}.
\bibitem{2h2} R.\ Atanasov, F.\ Bassani, and V.\ M.\ Agranovich, Second-order nonlinear optical susceptibility of asymmetric quantum wells, \href{https://doi.org/10.1103/PhysRevB.50.7809}{\prb\ \textbf{50}, 7809 (1994)}.
%\newpage
\bibitem{2h3} I.\ Paradisanos, A.\ M.\ S.\ Raven, T.\ Amand, C.\ Robert, P.\ Renucci, K.\ Watanabe, T.\ Taniguchi, I.\ C.\ Gerber, X.\ Marie, and B.\ Urbaszek, Second harmonic generation control in twisted bilayers of transition metal dichalcogenides, \href{https://doi.org/10.1103/PhysRevB.105.115420}{\prb\ \textbf{105}, 115420 (2022)}.
\bibitem{PGE1} E.\ L.\ Ivchenko and G.\ E.\ Pikus, New photogalvanic effect in gyrotropic crystals, Pis'ma\ Zh.\ Eksp.\ Teor.\ Fiz.\ \textbf{27}, 640 (1978); [JETP Lett. {\bf 27}, 604 (1978)].
\bibitem{PGE2} V.\ I.\ Belinicher,  Space-oscillating photocurrent in crystals without symmetry center, \href{https://doi.org/10.1016/0375-9601(78)90660-6}{Phys.\ Lett.\ A\ \textbf{66}, 213 (1978)}.
\bibitem{PGE3} S.\ D.\ Ganichev and L.\ E.\ Golub, Interplay of Rashba/Dresselhaus spin splittings probed by photogalvanic spectroscopy --- A review, \href{https://doi.org/10.1002/pssb.201350261}{Phys.\ Stat.\ Solidi\ (b)\ \textbf{251}, 1801 (2014)}.
\bibitem{NHE2} Z.\ Z.\ Du, H.-Z.\ Lu, and X.\ C.\ Xie, Nonlinear Hall effects, \href{https://doi.org/10.1038/s42254-021-00359-6}{Nat.\ Rev.\ Phys.\ \textbf{3}, 744 (2021)}.
\bibitem{NHE3} B.\ Cheng, Y.\ Gao, Z.\ Zheng, S.\ Chen, Z.\ Liu, L.\ Zhang, Q.\ Zhu, H.\ Li, L.\ Li, and C.\ Zeng, Giant nonlinear Hall and wireless rectification effects at room temperature in the elemental semiconductor tellurium, \href{https://doi.org/10.1038/s41467-024-49706-y}{Nat.\ Commun.\ \textbf{15}, 5513 (2024)}.
\bibitem{nye} J.\ F.\ Nye, \textit{Physical Properties of Crystals} (Oxford University Press, London, 1985). %ISBN:\ \href{https://isbnsearch.org/isbn/9780198511656}{9780198511656}
\bibitem{seldon1} A.\ C.\ Seldon, Analysis of the saturable absorber transmission equation, \href{https://doi.org/10.1088/0022-3727/3/12/323}{J.\ Phys.\ D: Appl.\ Phys.\ \textbf{3}, 1935 (1970)}.
\bibitem{seldon2} A.\ C.\ Seldon, Slow light and saturable absorption, \href{https://doi.org/10.1134/S0030400X09060150}{Opt.\ Spectrosc.\ \textbf{106}, 881 (2009)}.
\bibitem{Letokhov} V.\ S.\ Letokhov and V.\ P.\ Chebotayev, \textit{Nonlinear Laser Spectroscopy} (Springer, Berlin, 1977). %ISBN:\ \href{https://isbnsearch.org/isbn/9783662134870}{9783662134870}
\bibitem{rev} V.\ S.\ Zapasskii, Highly sensitive polarimetric techniques (review), \href{https://doi.org/10.1007/BF00663152}{J.\ Appl.\ Spectrosc.\ \textbf{37}, 857 (1982)}.
\bibitem{glaz} P.\ Glasenapp, A.\ Greilich, I.\ I.\ Ryzhov, V.\ S.\ Zapasskii, D.\ R.\ Yakovlev, G.\ G.\ Kozlov, and M.\ Bayer, Resources of polarimetric sensitivity in spin noise spectroscopy, \href{https://doi.org/10.1103/PhysRevB.88.165314}{\prb\ \textbf{88}, 165314 (2013)}.
\bibitem{depression} V.\ S.\ Zapasskii, Depression of excess light noise in polarimetric measurements, Opt.\ Spektrosk.\ \textbf{47}, 810 (1979); [Opt.\ Spectrosc.\ \textbf{47}, 450 (1979)]. 
\bibitem{optor} {\it Optical Orientation}, edited by F.\ Meier and B.\ Zakharchenya (North Holland, Amsterdam, 1984). %ISBN: \href{https://isbnsearch.org/isbn/9780444867414}{9780444867414}
\bibitem{cr4} Y.\ Sato and T.\ Taira, Model for the polarization dependence of the saturable absorption in Cr$^{4+}$:YAG, \href{https://doi.org/10.1364/OME.7.000577}{Opt.\ Mater.\ Express\ \textbf{7}, 577 (2017)}.
\bibitem{zscan} R.\ DeSalvo, M.\ Sheik-Bahae, A.\ A.\ Said, D.\ J.\ Hagan, and E.\ W.\ Van\ Stryland, Z-scan measurements of the anisotropy of nonlinear refraction and absorption in crystals, \href{https://doi.org/10.1364/OL.18.000194}{Opt.\ Lett.\ \textbf{18}, 194 (1993)}.
\bibitem{feofil} P.\ P.\ Feofilov and A.\ A.\ Kaplyanskii, Latent optical anisotropy of cubic crystals containing local centers and methods for its investigation, \href{https://doi.org/10.1070/PU1962v005n01ABEH003401}{Sov.\ Phys.\ Usp.\ \textbf{5}, 79 (1962)}.
\bibitem{budker3}D. Budker, D. F. Kimball, S. M. Rochester, V. V. Yashchuk, and M. Zolotorev, Sensitive magnetometry based on nonlinear magneto-optical rotation, \href{https://doi.org/10.1103/PhysRevA.62.043403}{\pra\ \textbf{62}, 043403 (2000)}. 
\bibitem{budker2}D. Budker and M. Romalis, Optical magnetometry, \href{https://doi.org/10.1038/nphys566}{Nat.\ Phys.\ \textbf{3}, 227 (2007)}.
\bibitem{jackson} D. F. Jackson Kimball, L. R. Jacome, S. Guttikonda, E. J. Bahr, and L. F. Chan, Magnetometric sensitivity optimization for nonlinear optical rotation with frequency-modulated light: Rubidium D2 line, \href{https://doi.org/10.1063/1.3225917}{J.\ Appl.\ Phys.\ \textbf{106}, 063113 (2009)}. 
\bibitem{makarova}A. O. Makarova, D. V. Brazhnikova, and A. N. Goncharova, Observation of the strong magneto-optical rotation of the polarization of light in rubidium vapor for applications in atomic magnetometry, \href{https://doi.org/10.1134/S0021364023600684}{JETP\ Lett.\ \textbf{117}, 509 (2023)}.

\end{thebibliography}
\end{document}